\documentclass[10pt]{article}
\usepackage{amsmath,amssymb,amsfonts,epsfig,graphics,graphicx}
\newcommand{\be}{\begin{equation}}
\newcommand{\ee}{\end{equation}}
\newcommand{\ea}{\end{eqnarray}}
\newcommand{\ba}{\begin{eqnarray}}

\newcommand{\rparen}{({\bf r})}

\newcommand{\bfp}{{\bf p}}

\newcommand{\sigvec}{\boldsymbol \sigma}
\newcommand{\gammavec}{{\boldsymbol \gamma}}

\newcommand{\alphavec}{{\boldsymbol \alpha}}
\newcommand{\Avec}{{\bf A}}

\begin{document}

\title{\Large A SUPERSYMMETRIC MODEL FOR GRAPHENE}

\author{Everton M. C. Abreu$^{1}$ \thanks{E-mail: evertonabreu@ufrrj.br}$\;$, Marco A. De Andrade
$^{2,5}$ \thanks{E-mail: marco@cbpf.br}$\;$, Leonardo P. G. de Assis$^{1,3}$ \thanks{E-mail: lpgassis@ufrrj.br}$\;$, \\
Jos\'{e} A. Helay\"{e}l-Neto$^{3,5}$ \thanks{E-mail: helayel@cbpf.br}$\;$,  
A. L. M. A. Nogueira$^{4,5}$ \thanks{E-mail: nogue@cbpf.br}$\;$, and Ricardo C. Paschoal$^{4,5}$ \thanks{E-mail: paschoal@cbpf.br}  \\  \\
$^{1}${\it  \normalsize Departamento de F\'{\i}sica, Universidade Federal Rural do Rio de Janeiro,} \\
{\it \normalsize BR 465-07, 23890-971, Serop\'edica, RJ, Brasil.}\\
$^{2}${\it \normalsize Universidade do Estado do Rio de Janeiro (Resende-RJ),} \\
{\it \normalsize Rodovia Presidente Dutra, km 298, P\'olo Industrial, CEP 27537-000, Resende, RJ, Brasil.}\\
$^{3}${\it  \normalsize Centro Brasileiro de Pesquisas F\'{\i}sicas -- CBPF,} \\
{\it \normalsize Rua Dr. Xavier Sigaud 150, 22290-180, Rio de Janeiro, RJ, Brasil.}  \\
$^{4}${\it  \normalsize Centro Federal de Educa\c{c}\~ao Tecnol\'ogica Celso Suckow
da Fonseca (CEFET/RJ), } \\
{\it \normalsize Av. Maracan\~{a}, 229, 20271-110, Rio de Janeiro, RJ, Brasil.} \\   
$^{5}${\it \normalsize Grupo de F\'{\i}sica Te\'{o}rica Jos\'e Leite Lopes,} \\
{\it \normalsize P.O.\ Box 91933, 25685-970, Petr\'opolis, RJ, Brasil.} \\         }

\date{\today}
\maketitle

\begin{abstract}
\noindent In this work, we focus on the fermionic structure of the low-energy
excitations of graphene (a monolayer of carbon atoms) to
propose a new supersymmetric field-theoretic model for this physical
system. In the current literature, other proposals for describing graphene
physics have been contemplated at the level of supersymmetric quantum
mechanics.
Also, by observing the inhomogeneities between neighbor carbon atoms,
Jackiw {\it
et al.} have set up an interesting chiral Abelian gauge theory.
We show in this paper that our formulation encompasses models discussed previously as
sectors
of an actually richer (supersymmetric) planar gauge model.
Possible interpretations for the fields involved in the present graphene model
are proposed and the question of supersymmetry breaking is discussed.
\end{abstract}

\pagestyle{myheadings}
\markright{{\it A Supersymmetry model for graphene}}

\section{Introduction}

Graphene is a flat, two-dimensional system, consisting of a monolayer of carbon atoms in a honeycomb array. Its electronic theoretical description is well known for decades \cite{Wallace,McClure,Sloncz}, and its analogy to planar quantum electrodynamics was used in the eighties \cite{Sem,Frad,Hal} for the study of certain aspects such as quantum anomalies. Its first experimental realization, however, occurred only in 2004 \cite{Nov1,Nov2}, with the follow-up experiments confirming the theory \cite{Nov3,Kim} and giving rise to a huge amount of work, a great part of which are summarized in various excellent introductory \cite{Int1,Int2,Int3,Int4,Int5} or more advanced \cite{Rev-1,Rev0,Rev1,Rev2,Rev3,Rev4,Rev5,Rev6,Rev7,Rev8} reviews already available. Thus, this (genuinely) planar carbon system appears to be a promising framework for the verification of ideas and methods developed in
quantum (gauge-) field theories, and therefore techniques of QED$_{3}$ may hopefully
yield a number of different and relevant results in this low-dimensional
condensed-matter system (see reviews \cite{RevRel1,RevRel2}). 

The fundamental electronic property of graphene \cite{Rev3} is the fact that its low-energy excitations are described by a Dirac Hamiltonian for massless particles (called Dirac fermions). So, at this level, the only field appearing in the theory is the Dirac spinor, which possesses four entries: two due to the fact that there are two atoms in each site of the triangular Bravais lattice (or, equivalently, there are two triangular sublattices performing the hexagonal one), and the remaining two due to the existence of two inequivalent points (the so-called `valleys', or Dirac points, which in literature are often named $\mathbf{K_+}$ and $\mathbf{K_-}$) of the first Brillouin zone. No considerations about the real spin of the electrons are being made here and for this reason such spinorial degrees of freedom in graphene carry a quantum number referred to as pseudospin~\footnote{Nevertheless, very recent works~\cite{Regan1,Regan2} present strong evidence (both experimental and theoretical) that pseudospin must be interpreted as a real angular momentum.}.

Another interesting issue related to graphene physics is the fact that an effective gauge field (Abelian or not) may also appear in the theory. This aspect is summarized in Refs.~\cite{Rev3,RevRel1} and in the specific reviews \cite{gauge1,gauge2}. Such an effective gauge field is induced if one or more of the following possibilities are taken into account (besides the obvious one corresponding to an  electromagnetic field): ripples \cite{gauge3} or curvature effects of the graphene sheet; lattice distortions or strains --- see, e.g., Ref.~\cite{strain}; hopping inhomogeneities between neighbour carbon atoms; or all of these, since one may induce the other(s). With respect to the third possibility, Jackiw and co-authors have \cite{jp,jp2,jp3} proposed a very interesting chiral (Abelian) gauge theory, in which also a complex-valued scalar field is present. The latter had been introduced previously by Hou, Chamon and Mudry \cite{hcm} and is also generated by the hopping inhomogeneities, more precisely, by the so-called Kekul\'e distortion, studied by Chamon in 2000~\cite{cham2000}. 

The presence of these three fields allows one to wonder if some relationship such as supersymmetry (SUSY) may exist among them and, eventually, other fields to be considered in addition (supersymmetric partners). Such a possibility is reinforced by some works \cite{Eza1,Eza2,Kailas,Kor,Chi,Ind} in which the possible manifestation of SUSY in a theory for graphene has already been raised, even though they remained at the level of supersymmetric quantum mechanics (and corresponding features of
the energy spectrum), still lacking a proposal for a (supersymmetric) graphene field theory.

Inspired by these results and ideas, our main contribution is to actually point out that the gauge theory constructed by Jackiw and collaborators \cite{jp,jp2,jp3} may be considered (at least in its original version of Ref.~\cite{jp}) as a sector of a wider and richer planar gauge
field theory: the so-called Supersymmetric $\tau_{3}$-QED \cite{DelCima-Marco}. With our work, supersymmetric quantum field theory becomes one more item to be added to the list \cite{RevRel1,RevRel2} of topics originated from high-energy physics which have the potentiality to be realized in table-top experiments with graphene. 

The present work is organized as follows: we shall review, in the next Section, some basic facts
about the mentioned chiral gauge theory and, in Section 3, we shall set up the main aspects of Supersymmetric $\tau_{3}$-QED. 
%\textbf{
Then, in Section 4, we shall explicitly build up a general
(power-counting renormalizable) supersymmetric action in (2+1)-dimensional space-time to finally show that
the action in Ref.~\cite{jp} (and possibly its generalizations in Refs.~\cite{jp2,jp3}) could be embodied in a broader supersymmetric functional. %}
Some conclusions are depicted in the last Section. A four-subsection Appendix is added with information about the representation of gamma matrices.

%%%%%%%%%%%%%%%%%%%%%%%%%%%%%%%%%%%%%%%%%%%%%%%%%%%%%%%%%%%%%%%%%%%%%%%%%%%%%%%%%
%

\section{A chiral gauge theory for graphene}

In 2000, Chamon studied in detail~\cite{cham2000} some consequences of the so-called Kekul\'e distortion in a honeycomb array of carbon atoms. In 2007, Hou, Chamon and Mudry (HCM) extended this idea~\cite{hcm} considering a Kekul\'e \emph{texture}, that is, a different Kekul\'e distortion in each point of the plane, thus introducing a scalar field, which will be represented here by $\varphi$. Some time later, in order to provide dynamics and finite energy to the vortices described by HCM, Jackiw and Pi~\cite{jp} introduced a gauge field $A_\mu$ to their model, thus formulating the chiral gauge theory that will be summarized in the present Section. For more details about this theory and its extensions~\cite{jp2,jp3}, the reader is referred to the original articles.

HCM's original theory for the Kekul\'e texture in graphene is described by the following Hamiltonian density~\cite{hcm}:
\begin{equation}
H =\int d^{2} r \Psi^{\dag}\, \rparen\, K \Psi\rparen,\label{rjeq6}%
\end{equation}
where $\Psi\rparen$ is a four-component spinor describing the electrons in graphene (called Dirac fermions, an expression meaning massless fermions)
\be
\Psi= \left(
\begin{array}
[c]{c}%
\psi^{b}_{+}\\[1ex]%
\psi^{a}_{+}\\[1ex]%
\psi^{a}_{-}\\[1ex]%
\psi^{b}_{-}%
\end{array}
\right), 
\ee
where the indices $a$ and $b$ refer to the two triangular sublattices of the honeycomb array and the indices $+$ and $-$ refer to the two inequivalent points (Dirac points) of the first Brillouin zone. 
$K$ is the $4\times4$ matrix \be\label{1} K = \left(
\begin{array}
[c]{cccc}%
0 & -2 i \partial_{z} & g\varphi\rparen & 0\\[1ex]%
-2 i \partial_{z} \ast & 0 & 0 & g\varphi\rparen\\[1ex]%
g\varphi^{\ast} \rparen & 0 & 0 & 2 i \partial_{z}\\[1ex]%
0 & g\varphi^{\ast}\rparen & 2 i \partial_{z}\ast & 0
\end{array}
\right),  \ee
with $-2 i \partial_{z} = \frac{1}{i}\ (\partial_{x} -i \, \partial_{y})$ and $g$ is a coupling constant. As mentioned above, the
complex-valued scalar $\varphi\rparen$ describes the Kekul\'e texture. Its modulus provides a
single-particle mass gap for the Dirac fermions and its phase may describe vortices.

Jackiw and Pi \cite{jp}, concerned with the fact that 
HCM's model leaves unspecified the dynamics that gives rise to the complex
vortex profile, proposed an extension of HCM's model
introducing a gauge potential and coupling it to the Dirac fermions in a chiral manner.

Using the Dirac matrices forms \be
\alphavec = (\alpha^{1}, \alpha^{2}, \alpha^{3}) = {\binom{\sigvec
\qquad0}{0\quad-\sigvec}} ~\qquad\beta= {\binom{0\quad I }{I\quad0}} \ee
and the chiral gamma matrix as the Hermitian one \be
\gamma_5^{J} = -i \alpha^{1}\, \alpha^{2}\, \alpha^{3} = {\binom{I\qquad0 }{0\quad
-I}}, \quad{(\gamma_5^{J}})^{2} = I , \ee
the gamma matrices are \be
\gammavec_J = \beta\, \alphavec = {\binom{0\quad-\sigvec }{
\sigvec\qquad0}} ,\quad\gamma^{0}_J = \beta,\quad\gamma_5^{J} = i\, \gamma
^{0}_J\, \gamma^{1}_J\, \gamma^{2}_J\, \gamma^{3}_J . \ee
With these matrices, $K$ in Eq.~(\ref{1}) may be represented as \be\label{2}
\Psi^{\dag}\, K \Psi= \Psi^{\dag}\ \left( \alphavec \cdot\bfp+ g \beta
\ [\varphi^{r} - i \varphi^{i}\, \gamma_5^{J} ] \right)\ \Psi, \ee
where $\bfp$ is the operator $-i\nabla$ and $\varphi\equiv\varphi^{r} +i\varphi^{i}$. 
Thus, the final Dirac Hamiltonian density, with the additional gauge potential, can be written as \ba
\Psi^{\dag}\, K_A \, \Psi & = &  \Psi^{\dag}\, \alphavec \cdot[\bfp-
q\gamma_5^{J}\, \Avec] \ \Psi+ g \Psi^{\dag}\beta\, [\varphi^{r} -
i\, \gamma_5^{J}\, \varphi^{i} ] \Psi\hspace{.72in}\nonumber\\[1ex] & = &
\bar{\Psi}_+\, \gammavec_J \cdot(\bfp- q\Avec) \ \Psi_+ +
\bar{\Psi}_- \, \gammavec_J \cdot(\bfp+ q\Avec) \Psi_- + g \,
\varphi\, \bar{\Psi}_+ \, \Psi_- + g \varphi^{\ast}\, \bar{\Psi}_- \, \Psi_+
\, , \label{rjeq11} \ea
where $q$ is a coupling constant describing the chiral charge (in Ref.~\cite{jp}, $q$ is set to unity), the Dirac adjoint is $\bar{\Psi} \equiv\Psi^{\dag}\, \gamma^{0}_{J}$ and
the chiral components are $\Psi_{\pm}\equiv\frac{1}{2} \ (1 \pm\gamma_{5}^{J})
\ \Psi$. The
Hamiltonian density which is present in Eq.~(\ref{rjeq11}) is invariant under a 
local chiral transformation
\begin{equation}
\varphi\to e^{2 i q \omega}\, \varphi\;, \qquad\Psi\to e^{i q \omega\gamma_{5}%
^{J}}\, \Psi\;, \qquad\Avec \rightarrow\Avec +
\nabla\,\omega\;;\label{rjeq9}
\end{equation}
\begin{equation}
\Psi_{\pm}\to e^{\pm iq \omega}\ \Psi_{\pm}, \qquad\bar{\Psi}_{\pm}\to\bar
{\Psi}_{\pm}\, e^{\mp iq \omega} .\label{rjeq12}
\end{equation}

Notice that this system possesses a global fermion number symmetry, with just
the Fermi fields transforming with a constant phase: $\Psi\to e^{i\lambda}\,
\Psi$. Consequently, the theory possesses a local chiral $U(1)$ symmetry and a
global $U(1)$ fermion number symmetry~\footnote{The gauged Abelian symmetry does {\it not} represent the usual interaction between electric charge and the gauge boson. As a matter of fact, it is the additional global Abelian phase symmetry that happens to be related to the electric charge.}. Because the theory resides in $(2+1)$
dimensions, no chiral anomalies interfere with the chiral gauge symmetry.

With the additional gauge potential $\Avec$, the Dirac eigenvalue problem in this model 
differs from HCM's. According to Eq.~(\ref{rjeq11}),
\begin{equation}
[\alphavec \cdot(\bfp-q\gamma_{5}^{J}\, \Avec) + g \beta(\varphi^{r} -i\,
\gamma_{5}^{J}\, \varphi^{i})]\Psi= E\, \Psi\, .\label{rjeq18}%
\end{equation}
Notice that $\alpha^{3}$, which will now be renamed as $R$, anti-commutes with the
matrix structure on the left side of Eq.~(\ref{rjeq18}). Therefore, if
$\Psi_{E}$ is an eigenfunction with eigenvalue $E$, $R\Psi_{E}$ belongs to
eigenvalue $- E$, and zero modes can be chosen as eigenstates of $R$. This is
a consequence of the ``sublattice symmetry" identified by HCM and renamed as energy-reflection symmetry in Ref.~\cite{jp3}.
%%%%%%%%%%%%%%%%%%%%%%%%%%%%%%%%%%%%%%%%%%%%%%%%%%%%%%%%%%%%%%%%%%%%%%%%%%%%%%%%%%%%%%%%%%%%%%%%%%%
%

From Eq.~(\ref{rjeq11}), we can write down the Lagrangian density
\begin{equation}
\mathcal{L}_{\mathrm{HCM-JP}}=\overline\Psi_{+}\gamma^{\mu}_{J}(iD_{\mu}^{+}%
)\Psi_{+} + \overline\Psi_{-}\gamma^{\mu}_{J}(iD_{\mu}^{-})\Psi_{-} -
g\varphi\overline\Psi_{+}\Psi_{-} - g\varphi^{\ast}\overline\Psi_{-}\Psi_{+}
\,,\label{J-P4x4}%
\end{equation}
where $iD_{\mu}^{+}=i\partial_{\mu}-qA_{\mu}~$ and $~iD_{\mu}^{-}%
=i\partial_{\mu}+qA_{\mu}$.

To set up this Jackiw-Pi's graphene theory in the context of 
supersymmetric $\tau_{3}$-QED, which will be presented in the next Section, it is convenient to re-write the Lagrangian
density, given in Eq.~(\ref{J-P4x4}), in terms of two-component spinor
notation. The four-component spinors $\Psi$, $\Psi_{+}$ and $\Psi_{-}$ can be
written in terms of two-component spinors $\psi_{+} = \left(
\begin{array}
[c]{c}%
\psi^{b}_{+}\\[1ex]%
\psi^{a}_{+}%
\end{array}
\right)  ~$ and $~ \psi^{\prime}_{-} = \left(
\begin{array}
[c]{c}%
\psi^{a}_{-}\\[1ex]%
\psi^{b}_{-}%
\end{array}
\right) ,~ $ so that
\begin{equation}%
\begin{array}
[c]{lcr}%
\Psi= \left(
\begin{array}
[c]{c}%
\psi_{+}\\[1ex]%
\psi^{\prime}_{-}%
\end{array}
\right) ,~ & \Psi_{+} = \left(
\begin{array}
[c]{c}%
\psi_{+}\\[1ex]%
0
\end{array}
\right) ,~ & \Psi_{-} = \left(
\begin{array}
[c]{c}%
0\\[1ex]%
\psi^{\prime}_{-}%
\end{array}
\right) .
\end{array}
\label{Psis}%
\end{equation}

We wish to keep the symbol $\psi_{-}$ (without ``prime'') for future spinor
redefinition. The $4\times4$ Dirac matrices can be expressed in terms of the
minimal $2\times2$ Dirac matrices as
\begin{equation}
\gamma^{\mu}_{J}= \left(
\begin{array}
[c]{lr}%
0 & \gamma^{\mu}\gamma^{0}\\
\gamma^{0}\gamma^{\mu} & 0
\end{array}
\right) \label{gammas}%
\end{equation}
with $\mu$ spanning the values (0,1,2) and $\gamma^{\mu}=(\sigma_{z},
i\sigma_{y}, -i\sigma_{x})$. In the sequel, using Eq.~(\ref{Psis}) and
Eq.~(\ref{gammas}), we can write down $\mathcal{L}_{\mathrm{HCM-JP}}$ in terms
of two-component spinors and minimal $2\times2$ Dirac matrices as
\begin{equation}
\mathcal{L}_{\mathrm{HCM-JP}}=\overline\psi_{+}\gamma^{\mu}(iD_{\mu}^{+})\psi_{+}
+ \overline\psi_{-}\gamma^{\mu}(iD_{\mu}^{-})\psi_{-} - g\varphi\overline
\psi_{+}\psi_{-} - g\varphi^{\ast}\overline\psi_{-}\psi_{+} ~,\label{LHCM2}%
\end{equation}
where
\begin{equation}
\psi_{-}\equiv\gamma^{0}\psi^{\prime}_{-}= \left(
\begin{array}
[c]{c}%
\psi^{a}_{-}\\[1ex]%
-\psi^{b}_{-}%
\end{array}
\right) ~.
\end{equation}

%%%%%%%%%%%%%%%%%%%%%%%%%%%%%%%%%%%%%%%%%%%%%%%%%%%%%%%%%%%%%%%%%%%%%%%%%%%%%%%%%%%%%%%%%%%%%%%%%%%%%%%%%%%%%%%%
%

\section{\bigskip The Supersymmetric $\tau_{3}$-QED Action}

With the purpose of formulating the supersymmetric $\tau_{3}$-QED action, we
refer to the work by Salam and Strathdee \cite{Salam}, where the superspace
and the superfields in (3+1) space-time dimensions were originally introduced.
Extending these ideas to the present case, the elements of the superfield are
parameterized as ($x^{\mu},\theta$), where $x^{\mu}$ are the coordinates of the
space-time and the fermionic coordinates, $\theta$, are Majorana spinors,
$\theta^{c} $=$\theta$. Now, we are ready to introduce the (simple)
supersymmetric $\tau_{3}$-QED formulation by means of the superfield formalism. As a
first step, we define the complex scalar N=1, D=(2+1)-superfields with opposed
$U(1)$-charges, $\Phi_{+}$ and $\Phi_{-}$, as
\begin{equation}
\Phi_{\pm}=A_{\pm}+\overline{\theta}\psi_{\pm}-\frac{1}{2}\overline{\theta
}\theta F_{\pm}~\text{ and }~\Phi_{\pm}^{\dagger}=A_{\pm}^{\ast}%
+\overline{\psi}_{\pm}\theta-\frac{1}{2}\overline{\theta}\theta F_{\pm}^{\ast}
~,\label{scalar3}%
\end{equation}
where $A_{\pm}$ are complex scalars, $\psi_{\pm}$ are Dirac spinors and
$F_{\pm}$ are auxiliary scalar fields that obey the following supersymmetric
transformations:
\begin{align}
\delta A_{\pm}  & = \overline{\varepsilon}\psi_{\pm}\\
\delta\psi_{\pm}  & = \varepsilon F_{\pm}+i\varepsilon\gamma^{\mu}%
\partial_{\mu}A_{\pm}\\
\delta F_{\pm}  & = i\,\overline{\varepsilon}\gamma^{\mu}\partial_{\mu}%
\psi_{\pm}~.
\end{align}

Notice that, by introducing $A_\pm$ and $\psi_\pm$ together in the same superfield, the scalars cannot be neutral under the global fermion number $U(1)$ symmetry~\cite{jp,jp3} mentioned in the previous Section.

In (2+1)-D, N=1-SUSY accommodates neutral matter in real scalar superfields. To describe charged fields, we have to combine two real supermultiplets. 

In the Wess-Zumino gauge, the gauge superconnection, $\Gamma_{a}$, is written
as
\begin{equation}
\Gamma_{a}=i\left(  \gamma^{\mu}\theta\right) _{a}A_{\mu}+\overline{\theta
}\theta\lambda_{a}~\text{ and }~\overline{\Gamma}_{a}=-i\left(  \overline
{\theta}\gamma^{\mu}\right) _{a}A_{\mu}+\overline{\theta} \theta
\overline{\lambda}_{a} ~,\label{gauge3}%
\end{equation}
where $A_{\mu}$ is the gauge boson and $\lambda_{a}$ is its partner, the gaugino (Majorana
spinor). Defining the ``field-strength'' superfield $W_{a}$ as
\begin{equation}
W_{a}=-\frac{1}{2}\overline{D}_{b}D_{a}\Gamma_{b}~,
\end{equation}
with covariant derivatives given by
\begin{equation}
D_{a}=\overline{\partial}_{a}-i\left(  \gamma^{\mu}\theta\right)  _{a}%
\partial_{\mu}~\text{ and }~\overline{D}_{a}=-\partial_{a}-i\left(
\overline{\theta}\gamma^{\mu}\right) _{a}\partial_{\mu}~,
\end{equation}
we obtain
\begin{equation}
W_{a}=\lambda_{a}+\Sigma^{\mu\nu}_{ab}\theta_{b}F_{\mu\nu}-\frac{i}
{2}\overline{\theta}\theta\gamma^{\mu}_{ab}\left( \partial_{\mu} \lambda
_{b}\right) \label{strenght3a}%
\end{equation}
and
\begin{equation}
\overline{W}_{a}=\overline{\lambda}_{a}-\overline{\theta}_{b}\Sigma^{\mu\nu
}_{ab}F_{\mu\nu}-\frac{i}{2}\overline{\theta}\theta\left( \partial_{\mu}
\overline{\lambda}_{b}\right) \gamma^{\mu}_{ab}~,\label{strenght3b}%
\end{equation}
where $\Sigma^{\mu\nu}=\frac{1}{4}\left[  \gamma^{\mu},\gamma^{\nu}\right]  $
are the generators of the Lorentz group in (2+1) space-time dimensions.

The gauge covariant derivatives that act on the matter fields with opposed
$U(1)$-charges, $\Phi_{+}$ and $\Phi_{-}$, are respectively given by
\begin{equation}
\nabla\Phi_{\pm}=\left(  D_{a}\mp iq\Gamma_{a}\right)  \Phi_{\pm}~\text{ and
}~\overline{\nabla}\Phi_{\pm}^{\dagger}=\left(  \overline{D}_{a}\pm
iq\overline{\Gamma}_{a}\right)  \Phi_{\pm}^{\dagger} ~.\label{deriv3}%
\end{equation}

Using the definitions previously given for the superfields in
Eqs.~(\ref{scalar3}), (\ref{gauge3}), (\ref{strenght3a}), (\ref{strenght3b})
and the gauge covariant derivatives given in Eq.~(\ref{deriv3}), we can build the supersymmetric $\tau_{3}$-QED action \cite{DelCima-Marco}:
\begin{equation}
S_{\tau_{3}\text{-QED}}= {\displaystyle\int} d^{3}xd^{2}\theta\left\{
-\frac{1}{2}\overline{W}W+(\overline{\nabla}\Phi_{+}^{\dagger})(\nabla\Phi
_{+})+(\overline{\nabla}\Phi_{-}^{\dagger})\left( \nabla\Phi_{-}\right)
+2m\left( \Phi_{+}^{\dagger}\Phi_{+}-\Phi_{-}^{\dagger}\Phi_{-}\right)
\right\}  ~.\label{superqedtau3}%
\end{equation}

In terms of component fields, the action given by Eq.~(\ref{superqedtau3}) may
be written as
\begin{align}
S_{\tau_{3}\text{-QED}} & = \int{d^{3}{x}}\left\{  {\frac
12}i{\overline\lambda}{\gamma^{\mu}{\partial}_{\mu}}\lambda-\frac14 F_{\mu\nu}%
F^{\mu\nu}+\right. \nonumber\\
&  - A_{+}^{\ast}\Box A_{+} - A_{-}^{\ast}\Box A_{-} + i \overline\psi_{+}
\gamma^{\mu}\partial_{\mu}\psi_{+} + i \overline\psi_{-} \gamma^{\mu}%
\partial_{\mu}\psi_{-} + F_{+}^{\ast}F_{+} + F_{-}^{\ast}F_{-} +\nonumber\\
&  - q A_{\mu}\biggr(\overline\psi_{+}\gamma^{\mu}\psi_{+} -\overline\psi
_{-}\gamma^{\mu}\psi_{-} + iA_{+}^{\ast}\partial^{\mu}A_{+} - iA_{-}^{\ast
}\partial^{\mu}A_{-} - iA_{+}\partial^{\mu}A_{+}^{\ast}+ iA_{-}\partial^{\mu
}A_{-}^{\ast}\biggr) +\nonumber\\
&  - i q \biggr(A_{+} \overline\psi_{+}\lambda- A_{-}\overline\psi_{-}\lambda-
A_{+}^{\ast}\overline\lambda\psi_{+} + A_{-}^{\ast}\overline\lambda\psi_{-}
\biggr) + q^2 A_{\mu}A^{\mu}\biggr(A_{+}^{\ast}A_{+} + A_{-}^{\ast}A_{-} \biggr
) +\nonumber\\
&  \left. -m\biggr(\overline\psi_{+}\psi_{+} - \overline\psi_{-}\psi_{-} +
A_{+}^{\ast}F_{+} - A_{-}^{\ast}F_{-} + A_{+}F_{+}^{\ast}-A_{-}F_{-}^{\ast}
\biggr) \right\} ~.\label{action3diag}%
\end{align}

As one can easily read from Eq.~(\ref{action3diag}), the two-flavoured fermionic sector presents  diagonal minimal coupling with the gauge boson, in correspondence with the first two terms displayed in Eq.~(\ref{LHCM2}). Nevertheless, the Yukawa-like interaction sector of Eq.~(\ref{LHCM2}) does not show up in the above $\tau_{3}$-supersymmetric  Lagrangian, even if we take the auxiliary fields $F_{+}$ and $F_{-}$ on-shell. As a matter of fact, a $\Phi^{4}$-sector has to be supplemented in order to provide the theory with the proper vertices, and such an extension must adopt the symmetry content of Lagrangian (\ref{action3diag}) as a guideline for its construction. Thus, we re-state the invariance of the supersymmetric $\tau_{3}$-Lagrangian with respect to an Abelian gauge transformation associated to the potential $A_{\mu}$ and to the field-strength $F_{\mu\nu}$, and we render evident a symmetry with respect to the following discrete {\it parity} variation:
\begin{eqnarray}
\Phi_\pm &\rightarrow & -\,\Phi_\mp \label{spparity}\, ,
\end{eqnarray}
as space-time transforms according to $(x^{0},x^{1},x^{2})\rightarrow(x^{0},- x^{1},x^{2})$.

The superfield variation expressed in Eq.~(\ref{spparity}) has its component-field counterpart spanned as
\begin{eqnarray}
\psi_\pm' &\equiv & \gamma^0\gamma^2 \psi_\mp \, ; \label{spinorpar} \\
A_{\pm}^{\prime} &\equiv & - A_{\mp} \, ; \nonumber \\
F_{\pm}^{\prime} &\equiv & F_{\mp} \, , \nonumber
\end{eqnarray}
where the 4-component spinor representation of parity reads
\begin{eqnarray}
\Psi^{\prime} &\equiv & {\left(\begin{array}{c}
\psi_{+} \\
\psi_{-}%\nonumber
 \end{array}\right)}^{\prime} \; = \; \Gamma_{\mbox{\tiny Par.}}\, \Psi \; = \; \left(\begin{array}{cc}
0 & \gamma^0\gamma^2 \\
\gamma^0\gamma^2 & 0 %\nonumber
 \end{array}\right) \; \Psi \, , %\nonumber
\end{eqnarray}
with $\Gamma_{\mbox{\tiny Par.}}$ defined in the context of a 4-spinor Pisarski representation, related to the Jackiw-Pi representation as drawn in the Appendix.

To complete the variation rule for the supersymmetric action, we shall comment on the fact that the Grassmannian (2+1)D-integration measure has an odd character with respect to parity,
\begin{equation}
d^{2}\theta\equiv d\overline{\theta}\,d\theta\stackrel{\mbox{\tiny Par.}}{\Longrightarrow} {d^{2}\theta}^{\prime} \, = \, -\, d^{2}\theta.
\end{equation}

This stems from imposing homogeneity on the effect of parity action onto superfields $\Phi_{\pm}$. The scalar superfields $\Phi_{\pm}$ accomodate a contracted two-component fermionic term in their expansions, namely, $\overline{\theta}\psi_{\pm}$. As $\psi_{\pm}$ transforms according to Eq.~(\ref{spinorpar}), $\overline{\theta}$ is required to change to\footnote{Such a variation also matches consistency as one considers both super-translations, $x^{\mu}\rightarrow x^{\mu} + i \overline{\epsilon}\gamma^{\mu}\theta$; $\theta \rightarrow \theta + \epsilon $, and parity, $x^{1}\rightarrow\, - \, x^{1}$; $x^{0}$,$x^{2}$ left unchanged. Also, the one-flavor content of the Grassmannian sector of superspace, as one deals with an N=1 theory, requires the variation of $\theta$ to be associated to a 2x2 block of $\Gamma_{\mbox{\tiny Par.}}$. As a room for further extensions, we comment on the fact that the model here proposed is obtained after a truncation of an N=2,D=(2+1) descent of a N=1,D=(2+2) supersymmetric theory \cite{DelCima-Marco}.} $\overline{\gamma^0\gamma^2\,\theta}$, up to a phase, as the term $({\overline{\theta}\psi_{\pm}})^{\prime}$ must still be a scalar. So, regardless the phase, $d^{2}\theta $ is necessarily odd under parity. As a consequence, the superpotential to be added to the theory has to be odd as well (note that the bare mass term proposed in Eq.~(\ref{superqedtau3}) already shares such a property).

\section{The $\Phi^{4}$-sector}

The most general local $U(1)$ and parity-invariant supersymmetric action with a $\Phi^{4}$-sector reads:
\begin{eqnarray}
S_{\Phi^{4}}&=& f {\displaystyle\int} d^{3}xd^{2}\theta\left[\left( \Phi_{+}^{\dagger}\Phi_{+}\right)^2 - \left(\Phi_{-}^{\dagger}\Phi_{-}\right)^2 \right] + \nonumber\\
 & & + h{\displaystyle\int} d^{3}xd^{2}\theta\left[ 
\left(\Phi_+^{\dag}\Phi_-\Phi_+\Phi_+ + \Phi_+\Phi_-^{\dag}\Phi_+^{\dag}\Phi_+^{\dag}\right) 
-\left( \Phi_-^{\dag}\Phi_+\Phi_-\Phi_- + \Phi_-\Phi_+^{\dag}\Phi_-^{\dag}\Phi_-^{\dag} \right)
  \right] \label{superphi4} \\
  & \equiv & S_f + S_h, \nonumber
\end{eqnarray}
where $f$ and $h$ are (real) coupling constants (to be associated later with Jackiw-Pi's $g$ present in Eq.~(\ref{LHCM2})). The quartic term with coupling constant $h$ explicitly breaks the global $U(1)$ fermion number symmetry. Nevertheless, we decide to keep this term in order to describe possible regimes where the chiral interaction dominates over the forces dictated by the global symmetry. The chiral symmetry has a dynamical character in that it dictates a gauge interaction; on the other hand, the global fermion number appears to have only a kinematic character. The latter classifies the states and field configurations (e.g., vortices in Refs.~\cite{jp,jp3}) without, however, introducing gauge-type interactions. Nothing prevents us from setting $h=0$ whenever we wish to recover results for which fermion number conservation is mandatory.

The component-wise $\Phi^{4}$-action expression follows:
\begin{equation*}
S_f \equiv {\displaystyle\int}d^{3}x {\cal L}_f,
\end{equation*}
with
\begin{eqnarray}
{\cal L}_f &=& 
-f\biggl[\left(F_+A^*_+ + F_+^*A_+\right)|A_+|^2 - \left(F_-A^*_- + F_-^*A_-\right)|A_-|^2 + 2\left(|A_+|^2\overline\psi_+\psi_+ - |A_-|^2\overline\psi_-\psi_-\right) + \nonumber \\
& & \left. +\frac{1}{2}\left( A_+^2\overline\psi_+\psi_+^c + A_+^{*2}\overline{\psi_+^c}\psi_+
- A_-^2\overline\psi_-\psi_-^c - A_-^{*2}\overline{\psi_-^c}\psi_-\right) \right] \label{gsector}.
\end{eqnarray}

Also, $S_h \equiv {\displaystyle\int}d^{3}x {\cal L}_h$ hosts the Lagrangian density
\begin{eqnarray}
-{\cal L}_h/h &=& {} + |A_+|^2\left(A_-F_+ + A_-^*F_+^* + \frac{1}{2}(A_+F_- + A_+^*F_-^*)
+ \overline{\psi_+^c}\psi_- + \overline\psi_-\psi_+^c \right) + \nonumber\\
& & {} - |A_-|^2\left(A_+F_- + A_+^*F_-^* + \frac{1}{2}(A_-F_+ + A_-^*F_+^*)
+ \overline{\psi_-^c}\psi_+ + \overline\psi_+\psi_-^c \right) + \nonumber\\
& & {} + \frac{1}{2}\left(A_+^2\overline\psi_+\psi_- + A_+^{*2}\overline\psi_-\psi_+
- A_-^2\overline\psi_-\psi_+ - A_-^{*2}\overline\psi_+\psi_- \right)+ \nonumber\\
& & {} + \left(A_-A_+ + A_-^*A_+^*\right)\left(\overline\psi_+\psi_+ - \overline\psi_-\psi_-\right)+ \nonumber\\
& & {} + \frac{1}{2}\left[A_+^*A_-\left(\overline{\psi_+^c}\psi_+-\overline\psi_-\psi_-^c\right)
+ A_-^*A_+\left(\overline\psi_+\psi_+^c-\overline{\psi_-^c}\psi_-\right)  \right] + \nonumber \\
& & +\frac{1}{2}\left( A_+^2A_-F_+^* + A_+^{*2}A_-^*F_+ 
- A_-^2A_+F_-^* - A_-^{*2}A_+^*F_-\right) \label{hsector}\, ,
\end{eqnarray}
where the two-component charge-conjugated spinors are defined by $\psi_{+}%
^{c}\equiv i\sigma_{x}\psi_{+}^{\ast}$ and $\psi_{-}^{c}\equiv i\sigma_{x}%
\psi_{-}^{\ast}$.

As we aim to end up with an action on-shell for the auxiliary sector, the field equations for $F_{\pm}$ are obtained:
\begin{eqnarray}
\frac{\delta{S}}{\delta{F_+}}&=&F_+^*-mA_+^*-f{A_+^*}|A_+|^2-h\left[|A_+|^2A_-+\frac{1}{2}\left( A_+^{*2}A_-^*-|A_-|^2A_- \right)\right]=0  \nonumber\\
 &&\nonumber\\
\frac{\delta{S}}{\delta{F_-^\ast}}&=&F_-+mA_-+f{A_-}|A_-|^2+h\left[|A_-|^2A_+^*+\frac{1}{2}\left( A_-^2A_+-|A_+|^2A_+^* \right)\right]=0  \nonumber\\
 &&\nonumber\\ \frac{\delta{S}}{\delta{F_+^\ast}}&=&F_+-mA_+-f{A_+}|A_+|^2-h\left[|A_+|^2A_-^*+\frac{1}{2}\left( A_+^2A_--|A_-|^2A_-^* \right)\right]=0  \nonumber\\
 &&\nonumber\\
\frac{\delta{S}}{\delta{F_-}}&=&F_-^*+mA_-^*+f{A_-^*}|A_-|^2+h\left[|A_-|^2A_++\frac{1}{2}\left( A_-^{*2}A_+^*-|A_+|^2A_+ \right)\right]=0. \label{Fs}
 \end{eqnarray}

The physical scalar sector of the complete action (that results from summing up Eq.~(\ref{superqedtau3}) and Eq.~(\ref{superphi4})) reads as follows: 
\begin{eqnarray}
V & = & F_{+}^{*}\, F_{+} \; +\; F_{-}^{*}\, F_{-} \, = \nonumber \\
&& = \, m^{2}\left( {|A_{+}|}^{2}\, + \,  {|A_{-}|}^{2}\right)\, + \, 2mf \left( {|A_{+}|}^{4}\, + \,  {|A_{-}|}^{4}\right)\, + \, mh \left( {|A_{+}|}^{2}\, + \,  {|A_{-}|}^{2}\right)\left( A_{+} A_{-} \, + \, A_{+}^{*}A_{-}^{*} \right)\nonumber \\
&& +\,  \frac{3fh}{2}\, \left( {|A_{+}|}^{4}\, + \, {|A_{-}|}^{4}\right)\left( A_{+} A_{-} \, + \, A_{+}^{*}A_{-}^{*} \right)\, - \, fh\, {|A_{+}|}^{2}{|A_{-}|}^{2}\left( A_{+} A_{-} \, + \, A_{+}^{*}A_{-}^{*} \right)\, + \nonumber \\
&& + \, f^{2}\left( {|A_{+}|}^{6}\, + \, {|A_{-}|}^{6}\right)\, + \, \frac{{h}^{2}}{4}\left( {|A_{+}|}^{2}\, + \, {|A_{-}|}^{2}\right)\left( A_{+}^{2} A_{-}^{2} \, + \, {A_{+}^{*}}^{2}{A_{-}^{*}}^{2} \right)\, + \nonumber \\
&& + \, \frac{{h}^{2}}{4}\, {|A_{+}|}^{2}{|A_{-}|}^{2}\left( {|A_{+}|}^{2}\, + \,  {|A_{-}|}^{2}\right)\, + \, \frac{{h}^{2}}{4}\left( {|A_{+}|}^{6}\, + \,  {|A_{-}|}^{6}\right)\, , \label{potential}
\end{eqnarray}
where a minus sign has been omitted to allow for the direct identification of the potential $V$.

At this point, we would like to reassess the procedure based on parity-symmetry we have followed to propose the $\Phi^{4}$-sector. As a matter of fact, the mass parameters present in extended models that follow original Jackiw-Pi's chiral gauge theory \cite{jp2,jp3} stand for v.e.v.'s of scalar fields, and a particular field is taken to be {\it odd} with respect to parity. Coherently, the corresponding mass parameter flips its sign as parity acts upon fields. If we had assumed the parameter $m$ displayed in Eq.~(\ref{superqedtau3}) to be odd as well, the parity transformation here proposed would not leave the action invariant, as it would also change by a sign. Instead, we take the viewpoint of considering the mass parameter as a fixed ``bare" mass, here included for the sake of generality, left unchanged by parity as the physical mass would get contributions from non-trivial minima of the power-counting renormalisable $A^{6}$-potential (mind Eq.~(\ref{potential})). The corresponding critical values of proper combinations of $A$-fields should inherit the odd-parity property proposed by Jackiw et al. Also, as far as the parameters $f$ and $h$ are concerned, those are taken invariant under parity as well, for the premise of their identification with v.e.v's of scalar fields would lead to a non-renormalisable $A^{7}$-potential, which we better avoid.

For the simple case $h=0$, the associated potential bears four minima:  
\begin{eqnarray}
& & (0,0),\;\text{trivial case} \\
& & \left( \sqrt{-\frac{m}{f}}, 0 \right) \\
& & \left( 0, \sqrt{-\frac{m}{f}} \right) \label{assimmaiszero}\\
& & \left( \sqrt{-\frac{m}{f}}, \sqrt{-\frac{m}{f}} \right)\, ,
\end{eqnarray}
where the entries refer to values of $|A_+|$ and $|A_-|$, respectively. It is important to mention that, in order that spontaneous symmetry breaking takes place, it is necessary that $f>0$ and $m<0$. Supersymmetry is preserved, since the potential in Eq.~(\ref{potential}) is zero for anyone of the minima above.

If we consider the mass matrix for the whole set of original fermionic fields ($\lambda$, ${\psi_{+}}$, ${\psi_{+}^{c}}$, ${\psi_{-}}$, ${\psi_{-}^{c}}$) and evaluate its eigenvalues on, say, the asymmetric (\ref{assimmaiszero}) minimum, we are left with the following results: two degenerate null outcomes; one $-\,m$ eingenvalue; and two masses given by $-\frac{m}{2}(1 \pm \sqrt{2-\frac{4q^2}{mf}})$ [notice that, since $q$ is real, $m<0$ and $f>0$, these masses are real]. So, massive Dirac fermions appear in graphene (as in Ref.~\cite{jp}) by means of the simplest self-interaction scenario. In addition, the corresponding eigenvectors are easily obtained and, by virtue of the breaking of gauge symmetry by the v.e.v. of the scalars, they obviously do not possess anymore a definite chiral charge~\footnote{Of course, although not explicitly mentioned therein, the same occurs with the Fermi fields $\Psi_{\pm}$ in Ref.~\cite{jp} whenever the scalar $\varphi$ acquires a non-trivial v.e.v.}. Qualitative similar results are obtained when the two other non-trivial minima are considered. Massive charged fermions in graphene, after supersymmetry breaking takes place through some mechanism (we do not address to this issue here), might combine into scalar bound states and their attractive interaction may reveal non-trivial and interesting effects of the doubly-charged scalars.

\section{Conclusion and Perspectives}

Back to the component-wise expression for the spinor-scalar 3- and 4-vertex sectors of the complete supersymmetric action, we write
\begin{eqnarray}
&& S_{\mbox{\tiny sp-sc int.}}\,=\, \nonumber \\
&=& {\displaystyle\int} d^{3}x \left\{ - i q \biggr(A_{+} \overline\psi_{+}\lambda- A_{-}\overline\psi_{-}\lambda-
A_{+}^{\ast}\overline\lambda\psi_{+} + A_{-}^{\ast}\overline\lambda\psi_{-}
\biggr) \, -  \left[ 2f {|A_{+}|}^{2} + h \left( A_{+}A_{-} + A_{+}^{*}A_{-}^{*} \right)\right] \, {\overline{\psi}}_{+}\psi_{+} \right.  \nonumber \\
&+& \left[ 2f {|A_{-}|}^{2} + h \left( A_{+}A_{-} + A_{+}^{*}A_{-}^{*} \right)  \right]{\overline{\psi}}_{-}\psi_{-} \, - \, \left[ \frac{f}{2} {A_{+}^{*}}^{2} + \frac{h}{2} A_{+}^{*}A_{-} \right] \overline{{\psi}_{+}^{c}}\psi_{+} \, - \, \left[    \frac{f}{2} A_{+}^{2} + \frac{h}{2} A_{+}A_{-}^{*}\right]{\overline{\psi}}_{+}\psi_{+}^{c}  \nonumber \\
&+& \, \left[ \frac{f}{2} {A_{-}^{*}}^{2} + \frac{h}{2} A_{-}^{*}A_{+} \right] \overline{{\psi}_{-}^{c}}\psi_{-} \, + \, \left[    \frac{f}{2} A_{-}^{2} + \frac{h}{2} A_{-}A_{+}^{*}\right]{\overline{\psi}}_{-}\psi_{-}^{c} \, - \, h\,|A_{+}|^{2}\left( \overline{{\psi}_{+}^{c}}\psi_{-} + {\overline{\psi}}_{-}\psi_{+}^{c} \right)  \nonumber \\ 
&+& \left. \, h\,|A_{-}|^{2}\left( \overline{{\psi}_{-}^{c}}\psi_{+} + {\overline{\psi}}_{+}\psi_{-}^{c} \right) \, - \frac{h}{2}\left( A_{+}^{2}  -  {A_{-}^{*}}^{2} \right) {\overline{\psi}}_{+}\psi_{-} \, - \frac{h}{2}\left( {A_{+}^{*}}^{2}  -  A_{-}^{2} \right) {\overline{\psi}}_{-}\psi_{+} \right\}. \label{interac} 
\end{eqnarray}

The content of the above Lagrangian density together with the fermionic minimal couplings with the gauge boson displayed in Eq.~(\ref{action3diag}) demonstrate that the supersymmetric $\tau_{3}$-QED theory given
in Eq.~(\ref{superqedtau3}) supplemented by a $\Phi^{4}$-term given in
Eq.~(\ref{superphi4}) provides a theoretical framework that extends Jackiw-Pi's original chiral gauge theory~\cite{jp}, represented here by Eq.~(\ref{LHCM2}). Furthermore, room is also available to allow for accommodating extended models \cite{jp2,jp3} in a supersymmetric scenario, with
the expectation that it may describe some of the physical features of
graphene. Of course, some identifications are necessary: for example, in a crude comparison, one would identify Jackiw-Pi's scalar field $\varphi$ with squares of the type $A_{+}^{2}$, $A_{-}^{2}$ (and complex conjugates) which arise from our quartic superfield action, Eq.~(\ref{superphi4}). To exactly identify terms and degrees of freedom one should establish the proper combinations of fermions that diagonalize the mass matrix (eigenvectors) upon a particular choice of scalar fields configuration that minimizes the potential. This has been done at the end of the previous Section, for the case $h=0$, which is the one to be considered for a comparison with Ref.~\cite{jp}, since, as already mentioned after Eq.~(\ref{superphi4}), the $h$-terms break the global $U(1)$ symmetry, related to electric charge. All the results obtained are compatible with Ref.~\cite{jp}.
%The study of critical points for the complete potential (Eq.~(\ref{potential})) as well as the analysis of mass matrix eigenvectors will soon be reported elsewhere.

We would like to stress --- and this is a non-trivial remark ---
that the supersymmetric $\tau_{3}$-QED related to the chiral gauge theory is
nothing but a planar descent of an (N=1)-supersymmetric Abelian gauge theory
in (2+2) space-time dimensions~\cite{DelCima-Marco}. It remains to be understood --- and, hopefully,
we shall also be clarifying this issue soon --- how to relate graphene physics to
the dimensional reduction of a gauge theory set up in Atiyah-Ward space. This point
sets out as a rich problem to be understood.

A comment on the gaugino is in order. It is an additional field with respect to the work of Ref.~\cite{jp}. 
To properly interpret the physical role of this (neutral) fermionic field, one has to consider that it mixes with the charged fermions whenever the complex scalars acquire non-trivial vacuum expectation values, as shown at the end of the previous Section. In this case, the $\lambda$-field itself does not describe a mass eigenstate and, therefore, it cannot be interpreted as a physical excitation. So, after choosing a particular vacuum, it is possible to diagonalize the mass matrix in order to identify the correct combinations of the fermionic fields that correspond to the mass eigenvalues. This procedure is outlined in the previous Section, for the case $h=0$. We could also check that SUSY is not broken (for $h=0$). To the best of our knowledge, these properties cannot yet be accommodated into the experimental results of graphene.

The physics of graphene has a very rich phenomenology. Supersymmetry may indeed open up a number of interesting issues in connection with the couplings between the fermionic and bosonic excitations that may arise in the description of this monolayer system. The spontaneous breaking of N=1-supersymmetry in (2+1)-D is less trivial than the same question in (3+1)-D, since N=1 does not exhibit a complex structure underneath, as planar N=2-supersymmetry does. So, we conclude this work with the perspective to reassess the problem of spontaneous supersymmetry breaking in (2+1)-D by adopting graphene as a physical system to apply our results. In particular, supersymmetry breaking should help us to understand and classify the rich structure of couplings in graphene.

\section*{Acknowledgements}

EMCA would like to thank the kindness and hospitality of Centro Brasileiro de Pesquisas F\' isicas (CBPF), where part of this work was accomplished.
This work is supported in part by Funda\c{c}\~ao de Amparo \`a Pesquisa do Estado do Rio de Janeiro (FAPERJ) and Conselho Nacional de Desenvolvimento Cient\' ifico e Tecnol\'ogico (CNPq), Brazilian Funding Agencies.

\appendix
%\section*{Appendix}
\setcounter{section}{0}
\setcounter{equation}{0}
%\Alph{section}
\section{Appendix}
\subsection{General Considerations}
\def\theequation{\Alph{section}.\arabic{equation}}

All the gamma matrices respect the anticommutation and Hermitization relations respectively given by
\begin{equation}
\{ \gamma^\mu~,~\gamma^\nu \}=2\eta^{\mu\nu},~~{\gamma^\mu}^\dag=\gamma^0\gamma^\mu\gamma^0,
\end{equation}
where $\eta^{\mu\nu}={\rm diag}(1,-1,-1)$. The other gamma matrices will always have the following properties:
\begin{equation}
{\gamma_5}^2=1,~~{\gamma_5}^\dag=\gamma_5~;~~~~~~{\gamma^3}^2=-1,~~{\gamma^3}^\dag=-\gamma^3~
\end{equation}
and the additional anticommutation relations given by
\begin{equation}
\{ \gamma^\mu~,~\gamma_5 \}=0~,\qquad\{ \gamma^\mu~,~\gamma^3 \}=0~,\qquad\{ \gamma^3~,~\gamma_5 \}=0~.
\end{equation}

\subsection{Jackiw-Pi representation of gamma matrices}

The $4\times4$ gamma matrices in Jackiw-Pi rep.\ \cite{jp}, which will be called $\gamma^\mu_J$, can be expressed in terms of gamma matrices in minimal $2\times2$ representation, which we will represent by $\gamma^\mu$, as
\begin{equation}
\gamma^\mu_J=
\left(\begin{array}{lr}
   0 & \gamma^\mu\gamma^0   \\
   \gamma^0\gamma^\mu  & 0  \\
\end{array}\right)     \label{Jgammas}
\end{equation}
with $\mu$ spanning only the values (0,1,2) and 
\begin{equation}
	\gamma^\mu=(\sigma_z, i\sigma_y, -i\sigma_x)\,.
\end{equation}
The additional matrices $\gamma_5^J$ and $\gamma^3_J$, are given by
\begin{equation}
\begin{array}{lr}
   \gamma_5^J=
   \left(\begin{array}{lr}
      I & 0   \\
      0 & -I 
   \end{array}\right),
&
   \gamma^3_J=i\gamma^0_J\gamma^1_J\gamma^2_J\gamma_5^J=
   \left(\begin{array}{lr}
      0 & -\gamma^0   \\
      \gamma^0  & 0  \\
   \end{array}\right).\label{Jg5g3}
\end{array}
\end{equation}
Since $\gamma^3_J$ is not associated to the space-time, it will behave as a second $\gamma^5$-type matrix, that is, it anticommutes with all the gamma matrices associated to the space-time.

\subsection{Pisarski representation of gamma matrices}

The $4\times4$ gamma matrices in Pisarski rep. \cite{Pisarski}, which will be called $\gamma^\mu_P$, can be expressed in terms of gamma matrices in minimal $2\times2$ representation, which we will represent by $\gamma^\mu$, as
\begin{equation}
\gamma^\mu_P=
\left(\begin{array}{lr}
    \gamma^\mu & 0   \\
    0  & -\gamma^\mu  \\
\end{array}\right)     \label{Pgammas}
\end{equation}
with $\mu$ spanning only the values (0,1,2) and
\begin{equation}
	\gamma^\mu=(\sigma_z, i\sigma_y, -i\sigma_x)\,.
\end{equation}
The additional matrices $\gamma_5^P$ and $\gamma^3_P$, are given by
\begin{equation}
\begin{array}{lr}
   \gamma_5^P=
   \left(\begin{array}{lr}
      0 & -I   \\
      -I & 0 
   \end{array}\right),
&
   \gamma^3_P=i\gamma^0_P\gamma^1_P\gamma^2_P\gamma_5^P=
   \left(\begin{array}{lr}
      0 & -I   \\
      I  & 0  \\
   \end{array}\right).\label{Pg5g3}
\end{array}
\end{equation}
Since $\gamma^3_P$ is not associated to the space-time, it will behave as a second $\gamma^5$-type matrix, that is, it anticommutes with all the gamma matrices associated to the space-time.

\subsection{The transformations that relate Jackiw-Pi. rep. to Pisarski rep.}

With the aid of the matrix $S=
   \left(\begin{array}{lr}
      I & 0   \\
      0 & \gamma^0 
   \end{array}\right)$,
we can write 
\begin{eqnarray}
\gamma^\mu_P&=&S\gamma^\mu_J\gamma^3_{J}S \label{transf-gms}\\
\gamma_5^P&=&S\gamma_5^J\gamma^3_{J}S \label{transf-g5}\\
\gamma^3_P&=&S\gamma^3_JS \label{transf-g3}~.
\end{eqnarray}

Eq.~(\ref{transf-g3}) follows from $\gamma^3_P$ and $\gamma^3_J$ definitions given, respectively, in Eqs.~(\ref{Pg5g3}) and (\ref{Jg5g3}) and, also, from the transformations given in Eqs.~(\ref{transf-gms}) and (\ref{transf-g5}). In the sequel we will write down the inverse transformations,
\begin{eqnarray}
\gamma^\mu_J&=&S\gamma^3_P\gamma^\mu_PS \label{itransf-gms}\\
\gamma_5^J&=&S\gamma^3_P\gamma_5^PS \label{itransf-g5}\\
\gamma^3_J&=&S\gamma^3_PS \label{itransf-g3}~.
\end{eqnarray}


\begin{thebibliography}{99}
\bibitem{Wallace} Wallace, P. R., %The band theory of graphite. 
\emph{Phys. Rev.} \textbf{71}, 622-–634 (1947).

\bibitem{McClure} McClure, J. W., %Diamagnetism of graphite. 
\emph{Phys. Rev.} \textbf{104}, 666-–671 (1956).

\bibitem{Sloncz} Slonczewski, J. C. and Weiss, P. R., %Band structure of graphite. 
\emph{Phys. Rev.} \textbf{109}, 272–-279 (1958).

\bibitem{Sem} Semenoff, G. W., %Condensed-matter simulation of a three-dimensional anomaly. 
\emph{Phys. Rev. Lett.} \textbf{53}, 2449–-2452 (1984).

\bibitem{Frad} Fradkin, E., %Critical behavior of disordered degenerate semiconductors. 
\emph{Phys. Rev. B} \textbf{33}, 3263–-3268 (1986).

\bibitem{Hal} Haldane, F. D. M., %Model for a quantum Hall eff ect without Landau levels: Condensed-matter realization of the 'parity anomaly'. 
\emph{Phys. Rev. Lett.} \textbf{61}, 2015–-2018 (1988).

\bibitem{Nov1} Novoselov, K. S. \textit{et al}., %Electric field effect in atomically thin carbon films. 
\emph{Science} \textbf{306}, 666–-669 (2004).

\bibitem{Nov2} Novoselov, K. S. \textit{et al}., %Two-dimensional atomic crystals. 
\emph{Proc. Natl Acad. Sci. USA} \textbf{102}, 10451-–10453 (2005).

\bibitem{Nov3} Novoselov, K. S. \textit{et al}., %Two-dimensional gas of massless Dirac fermions in graphene. 
\emph{Nature} \textbf{438}, 197–-200 (2005).

\bibitem{Kim} Zhang, Y., Tan, J. W., Stormer, H. L. and Kim, P., %Experimental observation of the quantum Hall effect and Berry's phase in graphene. 
\emph{Nature} \textbf{438}, 201–-204 (2005).

\bibitem{Int1} Castro Neto, A. H., Guinea, F. and Peres, N. M. R., \emph{Phys. World} \textbf{19}, 33 (2006).

\bibitem{Int2} Katsnelson, M. I., \emph{Mater. Today} \textbf{10}, 20 (2007).

\bibitem{Int3} Geim, A. K., and MacDonald, A. H., \emph{Phys. Today} \textbf{60}, 35 (2007). %Graphene: Exploring carbon flatland.

\bibitem{Int4} Peres, N. M. R., \emph{Eur. Phys. News} \textbf{40} (3), 17--20 (2008). %GRAPHENE: nEwPhySiCS in Two DimEnSionS

\bibitem{Int5} Geim, A. K. and Kim, P., \emph{Sci. Am.} \textbf{298}, 90--97 (April 2008).

\bibitem{Rev-1} Ando, T. \emph{Physica E} \textbf{40}, 213 (2007).

\bibitem{Rev0} Fal'ko, V. I. and Geim, A. K., \emph{Eur. Phys. J. Special Topics} \textbf{148}, 1--4 (2007).

\bibitem{Rev1} Geim, A. K. and Novoselov, K. S., \emph{Nature Mater.} \textbf{6}, 183 (2007). %The Rise of Graphene.

\bibitem{Rev2} Gusynin, V. P., Sharapov, S. G., and Carbotte, J. P., \emph{Int. J. Mod. Phys. B}  \textbf{21}, 4611--4658 (2007). %AC CONDUCTIVITY OF GRAPHENE: FROM TIGHT-BINDING MODEL TO 2 + 1-DIMENSIONAL QUANTUM ELECTRODYNAMICS.

\bibitem{Rev3} Castro Neto, A. H., Guinea, F., Peres, N. M. R., Novoselov, K. S., and Geim, A. K., \emph{Rev. Mod. Phys.} \textbf{81}, 109 (2009).

\bibitem{Rev4} Geim, A. K., \emph{Science} \textbf{324}, 1530--1534 (2009). %Graphene: Status and Prospects

\bibitem{Rev5} Peres, N. M. R., \emph{J. Phys.: Condens. Matter} \textbf{21}, 323201 (2009).

\bibitem{Rev6} Peres, N. M. R., and Ribeiro, R. M., \emph{New Journal of Physics} \textbf{11} 095002 (2009).

\bibitem{Rev7} Allen, M. J., Tung, V. C., and Kaner, R. B., \emph{Chem. Rev.} \textbf{110}, 132 (2010).

\bibitem{Rev8} Abergel, D. S. L., Apalkov, V., Berashevich, J., Ziegler, K., and Chakraborty, T., \emph{Properties of Graphene: A Theoretical Perspective}, arXiv:1003.0391 [cond-mat.mtrl-sci] (2010).

\bibitem{RevRel1} Katsnelson, M. I., and Novoselov, K. S., \emph{Solid State Comm.} \textbf{143}, 3--13 (2007). %Graphene: new bridge between condensed matter physics and quantum electrodynamics 

\bibitem{RevRel2} Shytov, A., Rudner, M., Gu, N., Katsnelson, M., and Levitov, L., \emph{Solid State Commun.} \textbf{149}, 1087 (2009). %Atomic collapse, Lorentz boosts, Klein scattering, and other quantum-relativistic phenomena in graphene

\bibitem{Regan1} Mecklenburg, M., and Regan, B. C., \emph{Spin and the Honeycomb Lattice: Lessons from Graphene}, arXiv:1003.3715 [cond-mat.mes-hall] (2010).

\bibitem{Regan2} Mecklenburg, M., Woo, J., and Regan, B. C., \emph{Electron-photon interactions in graphene}, arXiv:1003.4419 [cond-mat.mes-hall] (2010).

\bibitem{gauge1} Sasaki, K., and Saito, R., \emph{Prog. Theor. Phys. Suppl.} \textbf{176}, 253 (2008). %Pseudospin and Deformation-induced Gauge Field in Graphene

\bibitem{gauge2} Pachos, J. K., \emph{Contemp. Phys.} \textbf{50}, 375--389 (2009). %Manifestations of topological effects in graphene

\bibitem{gauge3} Guinea, F., Horovitz, B., and Le Doussal, P., \emph{Solid State Commun.} \textbf{149}, 1140--1143 (2009). %Gauge fields, ripples and wrinkles in graphene layers

\bibitem{strain} Le\'on, G., Prada, E., San-Jose, P., and Guinea, F., \emph{Phys. Rev. B} \textbf{81}, 161402(R) (2010). % \emph{Effects of strains and magnetic fields on electronic transport in suspended graphene}, arXiv:0906.5267 [cond-mat.mes-hall].

\bibitem{jp} Jackiw, R., and Pi, S.-Y., \emph{Phys. Rev. Lett.} \textbf{98}, 266402 (2007).

\bibitem{jp2} Chamon, C., Hou, C.-Y., Jackiw, R., Mudry, C., Pi, S.-Y., and Schnyder, A. P., \emph{Phys. Rev. Lett.} {\bf 100}, 110405 (2008). %, arXiv:0707.0293 [cond-mat.str-el].

\bibitem{jp3} Chamon, C., Hou, C.-Y., Jackiw, R., Mudry, C., Pi, S.-Y., and Semenoff, G., \emph{Phys. Rev. B} {\bf 77}, 235431 (2008). %, arXiv:0712.2439 [hep-th].

\bibitem{hcm} Hou, C.-Y., Chamon, C., and Mudry, C., \emph{Phys. Rev. Lett.}
\textbf{98}, 186809 (2007).

\bibitem{cham2000} Chamon, C., \emph{Phys. Rev. B} {\bf 62}, 2806 (2000). 

\bibitem{Eza1} Ezawa, M., \emph{Physica E} \textbf{40}, 269--272 (2007).

\bibitem{Eza2} Ezawa, M., \emph{Phys. Lett. A} \textbf{372}, 924--929 (2008).

\bibitem{Kailas} Kailasvuori, J., \emph{Europhys. Lett.} \textbf{87}, 47008 (2009). %, %\emph{Pedestrian index theorem \`{a} la Aharonov-Casher for bulk threshold modes in corrugated multilayer graphene}, arXiv:0904.3807 [cond-mat.mes-hall].

\bibitem{Kor} Park, K.-S., and Yi, K. S., \emph{J. Kor. Phys. Soc.} \textbf{50} (6), 1678--1682 (2007). %Supersymmetric quantum mechanics in graphene

\bibitem{Chi} Jia, W.-Z., and Wang, S.-J, \emph{Commun. Theor. Phys.} \textbf{50} (2), 335--340 (2008). % Supersymmetric quantum mechanics and SUSY dependent SU(2) symmetry

\bibitem{Ind} Sahoo, S., and Das, S., \emph{Indian J. Pure \& Appl. Phys.} \textbf{47} (3), 186--191 (2009). %Supersymmetric structure of fractional quantum Hall effect in graphene

\bibitem{DelCima-Marco} De Andrade, M. A., and Del Cima, O. M., \emph{Phys. Lett. B}
\textbf{347}, 95 (1995); and \emph{Int. J. Mod. Phys. A} \textbf{11}, 1367 (1996).

\bibitem{Salam} Salam, A., and Strathdee, J. S., \emph{Phys.\ Lett.\ B} \textbf{51}, 353 (1974); and \emph{Nucl. Phys. B} \textbf{76}, 477 (1974).

\bibitem{Pisarski} Pisarski, R. D., {\em Phys. Rev. D} {\bf 29}, 2423 (1984); and {\em Phys. Rev. D} {\bf 44}, 1866 (1991).

\end{thebibliography}
\end{document}